\documentclass{PoS}
\usepackage{amsmath}
\usepackage{subfigure}

\newcommand{\gev}{\, {\rm GeV}}
\newcommand{\mev}{\, {\rm MeV}}

\newcommand{\be}{\begin{equation}}
\newcommand{\ee}{\end{equation}}
\newcommand{\bea}{\begin{eqnarray}}
\newcommand{\eea}{\end{eqnarray}}

\title{Momentum dependence of kaon semileptonic form factors with $N_f=2+1+1$ Twisted Mass fermions}

\ShortTitle{Kaon semileptonic form factors with $N_f=2+1+1$ Twisted Mass fermions}

\author{ N. Carrasco$^{(a)}$, P. Lami$^{(a,b)}$, V. Lubicz$^{(a,b)}$, \speaker{L. Riggio}$^{(a)}$, S. Simula$^{(a)}$

\\

\it $^{(a)}$ INFN, Sezione di Roma Tre, Rome, Italy. Email: \email{carrasco@fis.uniroma3.it}, \email{lorenzo.riggio@gmail.com}, \email{simula@roma3.infn.it}

\it $^{(b)}$ Dipartimento di Matematica e Fisica, Universit\`a  Roma Tre, Rome, Italy.  Email: \email{lamipaolo@gmail.com}, \email{lubicz@fis.uniroma3.it}

\\

\bf{For the ETM Collaboration}
}       

\abstract{We present a lattice QCD determination of the vector and scalar form factors of the kaon semileptonic decay $K \to \pi \ell \nu$, which is relevant for the determination of the CKM matrix element $|V_{us}|$ from experimental data.
Our results are based on the gauge configurations produced by the European Twisted Mass Collaboration with $N_f = 2+1+1$ dynamical fermions.
We simulated at three different values of the lattice spacing and with pion masses as small as $210$ MeV.
Our estimate for the vector form factor at zero 4-momentum transfer is $f_+(0) = 0.9709 (46)$, where the uncertainty is both statistical and systematic.
By combining our result with the latest experimental value of $f_+(0)|V_{us}|$ we obtain $|V_{us}| = 0.2230 (11)$, which satisfies the unitarity constraint of the Standard Model at the permille level using the updated determination of $|V_{ud}|$ coming from superallowed nuclear $\beta$ decays.
We present also the momentum dependence of the vector and scalar form factors in the whole range of values of the squared 4-momentum transfer measured in $K_{\ell 3}$ decays, obtaining a good agreement with the experimental data.}

\FullConference{The 33nd International Symposium on Lattice Field Theory,\\
		14-18 July, 2015\\
		Kobe, Japan}

\begin{document}

\section{Introduction and simulation details}
\label{sec:intro}

In the Standard Model (SM) the relative strengths of the flavor-changing weak currents are parametrized by the Cabibbo-Kobayashi-Maskawa (CKM) matrix \cite{CKM}.
The accurate determination of its matrix elements is therefore crucial both for testing the SM and for searching new physics. 

In this contribution we present a new determination of the matrix element $|V_{us}|$ coming from the study of the semileptonic $K \to \pi \ell \nu$ decay on the lattice.
We use the ensembles of gauge configurations produced by the European Twisted Mass (ETM) Collaboration with four flavors of dynamical quarks ($N_f = 2+1+1$), which include in the sea, besides two light mass-degenerate quarks, also the strange and the charm quarks with masses close to their physical values \cite{ETMC}. 

The gauge ensembles and the simulations are the same adopted in Ref.~\cite{Carrasco:2014cwa} to determine the up, down, strange and charm quark masses, as well as in Ref.~\cite{Carrasco:2014poa} to determine the leptonic decay constants $f_K / f_\pi$, $f_D$ and $f_{D_s}$.
In particular we use three different values of the lattice spacing to allow for a controlled extrapolation to the continuum limit, the smallest spacing being equal to $\simeq 0.06$ fm, and different lattice volumes with simulated pion masses ranging from $\simeq 210$ to $\simeq 450 \mev$.
The physics described by our lattice simulations corresponds to the isospin symmetric limit of QCD, where $m_u = m_d$, assuming also zero quark electric charges.

At each lattice spacing different values of the light and strange quark masses, $m_\ell$ and $m_s$, have been considered to study the dependence of the form factor $f_+(0)$ on $m_\ell$ and to perform a small interpolation in $m_s$ using a simple quadratic spline.
For the physical values of $m_\ell$ and $m_s$ we use the ones determined in Ref.~\cite{Carrasco:2014cwa}. 
Valence quarks are simulated at different values of the spatial momentum using non-periodic boundary conditions \cite{Bedaque:2004kc,deDivitiis:2004kq,Guadagnoli:2005be}.
Both the spacelike and the timelike regions of the squared $4-$momentum transfer $q^2$ are covered.
The semileptonic vector and scalar form factors, $f_+(q^2)$ and $f_0(q^2)$, are extracted from suitable combinations of three-point correlation functions to study their dependence on $q^2$, the light-quark mass $m_\ell$ and the lattice spacing $a$.

Our estimate for the vector form factor at zero 4-momentum transfer is
 \be
     f_+(0) = 0.9709 ~ (45)_{stat} (9)_{syst} = 0.9709 ~ (46) ~ .
     \label{eq:results}
 \ee
By combining our result with the latest experimental value of $f_+(0)|V_{us}| = 0.2165 (4)$ from Ref.~\cite{Moulson:2014cra} we obtain $|V_{us}| = 0.2230 (11)$, which satisfies the unitarity constraint of the SM at the permille level using the updated determination of $|V_{ud}|$ coming from superallowed nuclear $\beta$ decays, i.e.~$|V_{ud}| = 0.97417 (21)$  \cite{Hardy:2014qxa}.

We present also the $q^2$-dependence of the vector and scalar form factors, $f_+(q^2)$ and $f_0(q^2)$, in the whole range of values of $q^2$ measured in $K_{\ell 3}$ decays.
We fit our results extrapolated to the physical point adopting the same dispersive parameterization \cite{Bernard:2006gy,Bernard:2009zm} used to describe the experimental data \cite{Antonelli:2010yf,Moulson:2014cra}. 
The dispersive fit depends on two parameters, $\Lambda_+$ and $C$, which represent respectively the slope of the vector form factor $f_+(q^2)$ at $q^2 = 0$ (in units of $M_\pi^2$) and the scalar form factor $f_0(q^2)$ at the (unphysical) Callan-Trieman (CT) point $q^2 = q_{CT}^2 \equiv M_K^2 - M_\pi^2$ \cite{Callan:1966hu}.
Our final results are
 \be
     \Lambda_+ = 24.2 ~ (1.2) \cdot 10^{-3} ~, \qquad \rm{log}(C) = 0.1998 ~(138) ~ ,
     \label{eq:results_dispersive}
 \ee
which compare positively with the latest experimental results \cite{Moulson:2014cra}
 \be
     \Lambda_+^{exp} = 25.75 ~ (36) \cdot 10^{-3} ~, \qquad \rm{log}(C)^{exp} = 0.1985 ~ (70) ~ .
     \label{eq:exp_results_dispersive}
 \ee

\section{Extraction of the semileptonic form factors}
\label{sec:ff}

The matrix elements of the strangeness changing vector current $\hat{V}_\mu$ between kaon and pion states decompose into two form factors, $f_+(q^2)$ and $f_-(q^2)$, as
 \be
    \label{eq:matrixelement_Vmu}
    \langle \hat{V}_\mu \rangle \equiv  \left<  \pi (p^\prime) | \hat{V}_{\mu} | K(p)  \right> = (p_\mu + p_\mu^\prime) f_+(q^2) + (p_\mu - p_\mu^\prime) f_-(q^2) 
 \ee
with $q_\mu \equiv p_\mu - p_\mu^\prime$.
The scalar form factor $f_0$ is defined as
 \be
    \label{eq:f0def}
    f_0(q^2) = f_+(q^2) + \frac{q^2}{M_K^2 - M_\pi^2} f_-(q^2)
  \ee
and therefore the relation $f_+(0) = f_0(0)$ is satisfied by construction.
The form factor $f_0(q^2)$ is proportional to the 4-divergence of $\langle \hat{V}_\mu \rangle$, which in turn, thanks to the vector Ward-Takahashi identity, is related to the matrix element of the strangeness changing scalar density $\hat{S}$ between kaon and pion states, leading to
 \be
      \label{eq:matrixelement_S}
      \langle \hat{S} \rangle \equiv \langle  \pi(p_\pi) | \hat{S} | K(p_K)  \rangle = \frac{M_K^2 - M_\pi^2}{m_s - m_\ell} f_0(q^2) ~ .
 \ee
Equations (\ref{eq:matrixelement_Vmu}) and (\ref{eq:matrixelement_S}) represent a system of redundant relations between the two form factors $f_+(q^2)$ and $f_0(q^2)$ and the matrix elements $\langle \hat{V}_\mu \rangle$ and $\langle  \hat{S} \rangle$.
The latter can be extracted from the (Euclidean) time dependence of suitable combinations of three-point correlation functions connecting the initial and final pseudoscalar mesons through either the (bare) local vector current $V_\mu$ or the (bare) scalar density $S$.
Indeed we calculate the following quantities 
 \bea
      \label{eq:RVmu}
      R_\mu^{(V)}(t; \vec{p}_K, \vec{p}_\pi) & \equiv & 4 p_{K \mu} p_{\pi \mu} ~ \frac{C_{V_\mu}^{K \pi}(t, \frac{T}{2}; \vec{p}_K, \vec{p}_\pi) ~ 
      C_{V_\mu}^{\pi K}(t, \frac{T}{2}; \vec{p}_\pi, \vec{p}_K)}{C_{V_\mu}^{\pi \pi}(t, \frac{T}{2};  \vec{p}_\pi, \vec{p}_\pi) ~ 
      C_{V_\mu}^{K K}(t, \frac{T}{2}; \vec{p}_K, \vec{p}_K)} ~ , \\
      \label{eq:RS}
      R^{(S)}(t; \vec{p}_K, \vec{p}_\pi) & \equiv & R_0^{(V)}(t; \vec{0}, \vec{0}) ~ \overline{K}  ~ \frac{C_S^{K \pi}(t, \frac{T}{2}; \vec{p}_K, \vec{p}_\pi) ~ 
      C_S^{\pi K}(t, \frac{T}{2};  \vec{p}_\pi, \vec{p}_K)} {C_S^{K \pi}(t, \frac{T}{2}; \vec{0}, \vec{0}) ~ C_S^{\pi K}(t, \frac{T}{2};  \vec{0}, \vec{0})} ~ ,
 \eea
where $\overline{K}$ is a simple factor depending on meson and quark masses, and $t$ is the time distance between the insertion of the local currents and the source\footnote{The time distance between the source and the sink is kept fixed at half of the time extension of the lattice $T/2$.}.
Notice that Eqs.~(\ref{eq:RVmu})-(\ref{eq:RS}) do not depend on any renormalization constant and involve only three-point correlation functions.
In Eq.~(\ref{eq:RS}) the ratio $R_0^{(V)}(t; \vec{0}, \vec{0})$ is introduced to take advantage of the high-precision determination of the scalar form factor $f_0$ at kinematical end-point $q^2 = q_{max}^2 \equiv (M_K - M_\pi)^2$ via the time component of the vector current \cite{Becirevic:2004ya}, namely 
 \be
     \label{eq:f0qmax}
     R_0^{(V)}(t; \vec{0}, \vec{0}) _{ ~ \overrightarrow{t \gg a, ~ (T/2 - t) \gg a} ~ } [(M_K + M_\pi) f_0(q_{max}^2)]^2 ~ .
 \ee
The use of Eq.~(\ref{eq:RS}) improves significantly the statistical precision of the extracted matrix element $\langle \hat{S} \rangle$ with respect to the standard strategy based on the ratio between three-point and two-point correlation functions.

At large time distances one has
 \bea
      \label{eq:RVmu_larget}
      R_\mu^{(V)}(t; \vec{p}_K, \vec{p}_\pi) _{ ~ \overrightarrow{t \gg a, ~ (T/2 - t) \gg a} ~ } & & |\langle \pi(p_\pi) | \hat{V}_\mu | K(p_K) \rangle |^2 ~ , \\
      \label{eq:RS_larget}
      R^{(S)}(t; \vec{p}_K, \vec{p}_\pi) _{ ~ \overrightarrow{t \gg a, ~ (T/2 - t) \gg a} ~ } & & |\langle \pi(p_\pi) | \hat{S} | K(p_K) \rangle |^2 ~ ,     
  \eea
and therefore the form factor $f_+(q^2)$ and $f_0(q^2)$ can be determined by minimizing the $\chi^2$-variable constructed using all the extracted matrix elements $\langle \hat{V}_\mu \rangle$ and $\langle \hat{S} \rangle$.
The use of the information contained both in $\langle \hat{V}_\mu \rangle$ and in $\langle \hat{S} \rangle$ helps in increasing the precision of the extracted vector and scalar form factors with respect to the strategy of adopting only the information contained in $\langle \hat{V}_\mu \rangle$, which was previously adopted in Ref.~\cite{Carrasco:2014pta}.

After a small interpolation of our lattice data to the physical value of the strange quark mass $m_s = 99.6 (4.3) \mev$ determined in Ref.~\cite{Carrasco:2014cwa}, we present in the next Section the extrapolation of our results for the form factors $f_+(q^2)$ and $f_0(q^2)$ at the physical point and in the continuum limit.

\section{Combined chiral and continuum extrapolations}
\label{sec:fits}

In this Section we extrapolate the form factors $f_+(q^2)$ and $f_0(q^2)$ to the physical point in a wide range of values of $q^2$, which includes the $q^2$-region accessible to experiments, i.e.~from $q^2 = 0$ to the kinematical end-point $q^2 = q_{max}^2 = (M_K - M_\pi)^2 \to 0.129 \gev^2$. 

We perform a combined fit of the $q^2$-, $m_\ell$- and $a^2$-dependencies of the form factors using either the prediction inspired SU(2) ChPT proposed in Ref.~\cite{Lubicz:2010bv} or a modified $z$-expansion.
In both cases we include in our analysis the constraint arising from the CT theorem \cite{Callan:1966hu}, which relates in the SU(2) chiral limit the scalar form factor $f_0(q^2)$ calculated at the unphysical point $q^2 = q_{CT}^2 = M_K^2 - M_\pi^2$ to the ratio of the leptonic decay constants $f_K / f_\pi$.

Following Ref.~\cite{Lubicz:2010bv} the first Ansatz can be derived from the next-to-leading (NLO) SU(3) ChPT predictions for the kaon and pion loop contributions to the form factors \cite{Gasser:1984ux,Gasser:1984gg}, by performing an expansion in powers of the variable $x \equiv M_\pi^2 / M_K^2$ keeping only the ${\cal{O}}(x)$, ${\cal{O}}(x \log x)$ and ${\cal{O}}(\log(1 - s))$ terms, where $s \equiv q^2 / M_K^2$.
Moreover, taking into account that a simple pole Ansatz is able to reproduce quite well our lattice data for each gauge ensemble, and that the extracted slopes exhibit an almost linear dependence on the light-quark mass $m_\ell$ and on the squared lattice spacing $a^2$, our first Ansatz has the following form
\be
   \label{eq:ff_SU2}
   f_{+,0}(q^2) = \frac{ f_+^{SU(2)}(0) - \left[ x T_{+,0}^1(s) + T_{+,0}^2(s) \right] M_K^2 / (4 \pi f)^2} {1 - q^2 (1 + P_{+,0} M_\pi^2 + D_{+,0} a^2 + K_{+,0}^{FSE}) / 
                         M_{V,S}^2} \left( 1 + A_{+,0} q^2 + D a^2 \right) ~ , 
 \ee      
where the functions $T_{+,0}^{1,2}(s)$ are explicitly given in Ref.~\cite{Lubicz:2010bv} and, according to the SU(2) ChPT expansion \cite{Flynn:2008tg}, $f_+^{SU(2)}(0)$ can be written as
 \be
     \label{eq:f0_SU2}
     f_+^{SU(2)}(0) = F_+ \left[ 1 - \frac{3}{4} \xi \mbox{log}(\xi) + C_1 \xi + C_2 \xi^2 \right]
  \ee
with $\xi \equiv M_\pi^2 / (4 \pi f)^2$.
In Eq.~(\ref{eq:ff_SU2}) the quantity $K_{+,0}^{FSE}$ is a phenomenological term that takes into account the finite volume effects observed in the form factor slopes by comparing the data for the ensembles A40.24 and A40.32, which differ only by the lattice size (see Ref.~\protect\cite{Carrasco:2014cwa}), while $M_V$($M_S$) is the mass of the low-lying vector (scalar) resonance.
The value of $M_V$ is fixed at the mass of the $K^*(892)$ vector resonance, while $M_S$ is left as a free parameter.

The second Ansatz is a modified version of the $z$-expansion of Ref.~\cite{Bourrely:2008za}, namely
 \be
     \label{eq:zexpansion}
     f_{+,0}(q^2 ) = \frac{f_+^{SU(3)}(0) + \tilde{A}_{+,0} \left[ z - z_0 + (z^2 - z_0^2) / 2 \right] + \tilde{D} a^2 \left( M_K^2 - M_\pi^2 \right)^2} {1 - q^2 (1 + \tilde{P}_{+,0} M_\pi^2 + 
                             \tilde{D}_{+,0} a^2 + \tilde{K}_{+,0}^{FSE}) / \tilde{M}_{V,S}^2} ~ ,
 \ee
where $z = (\sqrt {t_+ - q^2}  - \sqrt {t_+ - t_0}) / (\sqrt {t_+ - q^2 } + \sqrt {t_+ - t_0})$, $t_0 \equiv \left( M_K  + M_\pi \right) \left( \sqrt {M_K}  - \sqrt {M_\pi} \right)^2$, $t_+ \equiv \left( M_K  + M_\pi \right)^2$ and $z_0 \equiv z(q^2 = 0)$.
In Eq.~(\ref{eq:zexpansion}), according to the SU(3) ChPT expansion \cite{Gasser:1984ux,Gasser:1984gg}, $f_+^{SU(3)}(0)$ can be written as
 \be
     \label{eq:f0_SU3}
     f_+^{SU(3)}(0) = 1 + f_2 + \Delta f ~ ,
  \ee
where $f_2$ is the NLO term, which does not depend on any low-energy constant and it is calculable in terms of meson masses \cite{Gasser:1984ux,Gasser:1984gg}, while the quantity $\Delta f$ represents higher order corrections.
According to the Ademollo-Gatto theorem \cite{Ademollo:1964sr} we parametrize the latter as 
 \be
    \label{eq:Deltaf}
    \Delta f = \left( {M_K^2  - M_\pi^2} \right)^2 \left[ {\Delta_0  + \Delta _1 M_\pi^2} \right] ~ .
 \ee

\begin{figure}[htb!]
\begin{center}
{\hspace*{1.25cm} \includegraphics[scale=0.5]{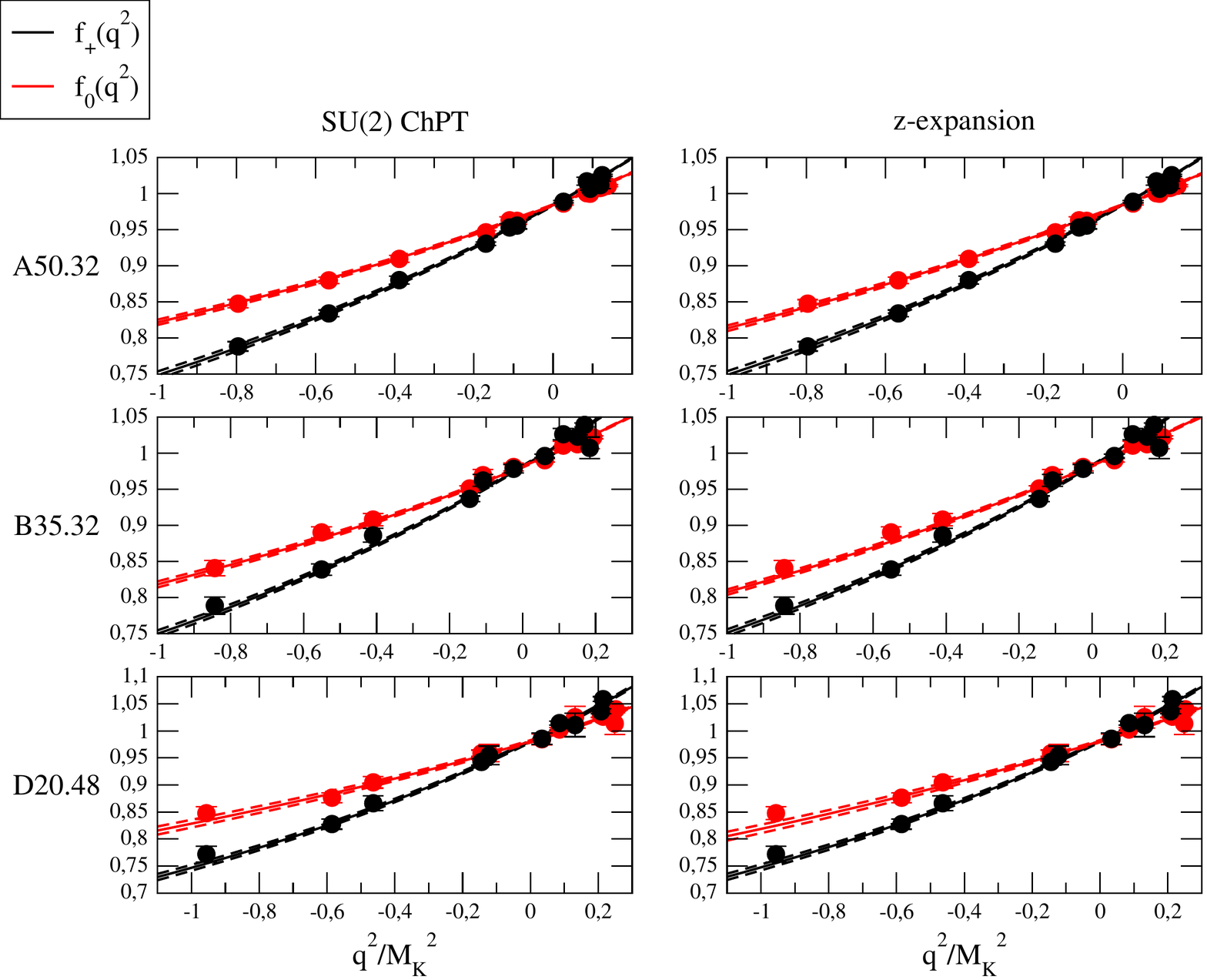}}
\end{center}
\vspace*{-1cm}
\caption{Lattice data for the vector and scalar form factors versus $q^2 / M_K^2$ for the ETMC ensembles A50.32, B35.32 and D20.48, corresponding to three values of the lattice spacing (see Ref.~\protect\cite{Carrasco:2014cwa}), compared with the results of the global fitting procedures based on Eq.~(\protect\ref{eq:ff_SU2}) (left panel) and Eq.~(\protect\ref{eq:zexpansion}) (right panel), shown as solid lines. The dashed lines represents the uncertainties of the fits.}
\label{fig:f0fp_global_fits}
\end{figure}

The quality of the global fitting procedures, based on Eq.~(\ref{eq:ff_SU2}) for the SU(2) ChPT Ansatz and Eq.~(\ref{eq:zexpansion}) for the modified z-expansion one, is illustrated in Fig.~\ref{fig:f0fp_global_fits}.

\begin{figure}[htb!]
\vspace{-0.5cm}
\begin{center}
{\hspace*{0.5cm} \includegraphics[scale=0.5]{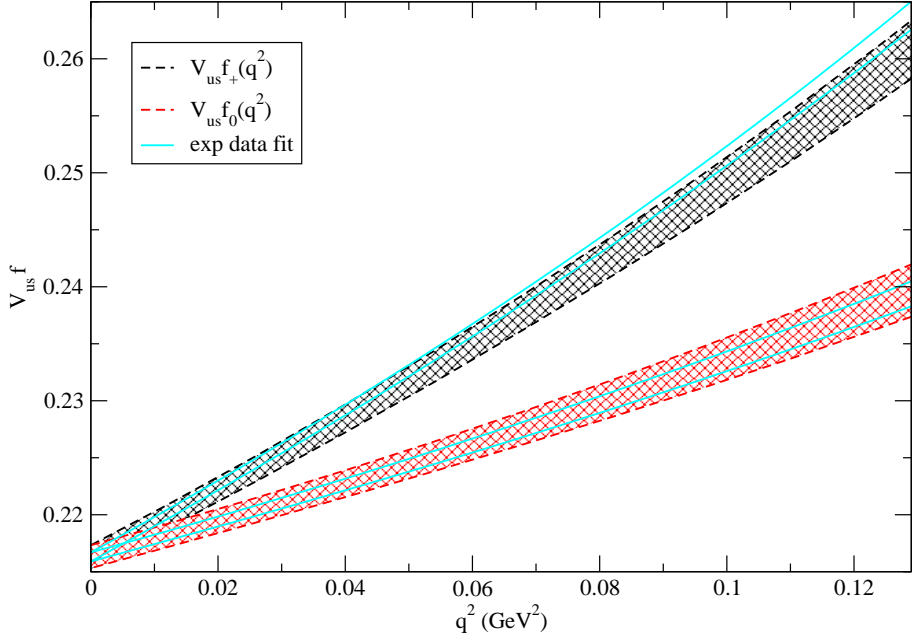}}
\end{center}
\vspace*{-1.25cm}
\caption{Results for the vector (black area) and scalar (red area) form factors, obtained at the physical point including both statistical and systematic uncertainties and multiplied by $|V_{us}| = 0.2230$, versus $q^2$. The cyan solid lines represent the results of the dispersive fit of the experimental data from Ref.~\cite{Moulson:2014cra}.}
\label{fig:f0fp_physical}
\end{figure}

Our fitting Ans\"atze are then used to extrapolate the form factors to the physical point and in the continuum limit.
Our findings compares positively with the results of the dispersive fit of the experimental data performed in Ref.~\cite{Moulson:2014cra}, as shown in Fig.~\ref{fig:f0fp_physical}.

Combining all the analyses we get the results reported in Eqs.~(\ref{eq:results})-(\ref{eq:results_dispersive}) for the form factor $f_+(0)$ and for the dispersive parameters $\Lambda_+$ and $\mbox{log}(C)$.
In Eqs.~(\ref{eq:results})-(\ref{eq:results_dispersive}) the statistical uncertainty includes also the error induced by the fitting procedure and the uncertainties of all the input parameters needed for the analyses, namely the physical values of the light and strange quark masses and the lattice spacing, determined in Ref.~\cite{Carrasco:2014cwa}.
The systematic error includes the uncertainties related to the chiral extrapolation, the discretization and finite volume effects, which will be discussed in details in a forthcoming publication \cite{ETMC_Kl3}.

\section{Calculation of $|V_{us}|$}
\label{sec:Vus}

Combining our result (\ref{eq:results}) with the updated experimental value of $f_+(0) |V_{us}| = 0.2165 (4)$ from Ref.~\cite{Moulson:2014cra} we can estimate the CKM matrix element $|V_{us}|$ obtaining
 \be
    \label{eq:Vus}
    |V_{us}| = 0.2230 ~ (4)_{exp} ~ (11)_{f_+(0)} = 0.2230 ~ (11) ~ , 
 \ee 
which can be compared with the determination $|V_{us}| = 0.2271 (29)$ obtained from $K_{\ell 2}$ decays using the ratio of the leptonic decay constants $f_{K^+} /f_{\pi^+}$ determined in Ref.~\cite{Carrasco:2014poa}.

Using our result (\ref{eq:Vus}) and taking the values $|V_{ud}| = 0.97417 (21)$ from the superallowed nuclear $\beta-$decays \cite{Hardy:2014qxa} and $|V_{ub}| = 4.13 (49) \cdot 10^{-3}$ from the PDG \cite{PDG}, we can test the unitarity of the first-row of the CKM matrix obtaining
 \bea
    \label{eq:utest}
     |V_{ud}|^2 + |V_{us}|^2 + |V_{ub}|^2 & = & 0.9988 ~ (9) \hspace*{1.2cm} \rm {from } ~~K_{\ell 3} ~~~\rm{[this~ work]} ~ , \nonumber \\
     |V_{ud}|^2 + |V_{us}|^2 + |V_{ub}|^2 & = & 1.0008 ~ (14) \hspace*{1cm} \rm {from } ~~K_{\ell 2} ~~~\mbox{\cite{Carrasco:2014poa}} ,
 \eea
which both confirm the SM constraint at the permille level.

\section*{Acknowledgements}

\noindent We acknowledge the CPU time provided by the PRACE Research Infrastructure under the project PRA027 ``QCD Simulations for Flavor Physics in the Standard Model and Beyond'' awarded at JSC (Germany), and by the agreement between INFN (Italy) and CINECA (Italy) under the specific initiative INFN-LQCD123.
V.~L., S.~S.~and C.~T.~thank MIUR (Italy) for partial support under Contract No. PRIN 2010-2011.
L.~R.~thanks INFN (Italy) for the support under the SUMA computing project (https://web2.infn.it/SUMA).


\begin{thebibliography}{99}

\bibitem{CKM} 
 N. Cabibbo, 
 Phys.\ Rev.\ Lett.\  {\bf 10} (1963) 531.

 M. Kobayashi and T. Maskawa, 
 Prog.\ Theor.\ Phys.\ {\bf 49} (1973) 652.

\bibitem{ETMC}
  R.~Baron {\it et al.} [ETM Coll.],
  JHEP {\bf 1006} (2010) 111
  [arXiv:1004.5284].

  R.~Baron {\it et al.}  [ETM Coll.],
  PoS LATTICE {\bf 2010} (2010) 123
  [arXiv:1101.0518].

 \bibitem{Carrasco:2014cwa}
  N.~Carrasco {\it et al.} [ETM Coll.],
  Nucl.\ Phys.\ B {\bf 887} (2014) 19
  [arXiv:1403.4504].

\bibitem{Carrasco:2014poa}
  N.~Carrasco {\it et al.} [ETM Coll.],
  Phys.\ Rev.\ D {\bf 91} (2015) 5, 054507
  [arXiv:1411.7908].
 
\bibitem{Bedaque:2004kc}
  P.~F.~Bedaque,
  Phys.\ Lett.\ B {\bf 593} (2004) 82
  [nucl-th/0402051].
  
\bibitem{deDivitiis:2004kq}
  G.~M.~de Divitiis, R.~Petronzio and N.~Tantalo,
  Phys.\ Lett.\ B {\bf 595} (2004) 408
  [hep-lat/0405002].
  
\bibitem{Guadagnoli:2005be}
  D.~Guadagnoli, F.~Mescia and S.~Simula,
  Phys.\ Rev.\ D {\bf 73} (2006) 114504
  [hep-lat/0512020].
 
  \bibitem{Moulson:2014cra}
  M.~Moulson,
  arXiv:1411.5252.

\bibitem{Hardy:2014qxa}
  J.~C.~Hardy and I.~S.~Towner,
  Phys.\ Rev.\ C {\bf 91} (2015) 2,  025501
  [arXiv:1411.5987].

\bibitem{Bernard:2006gy}
  V.~Bernard {\it et al.},
  Phys.\ Lett.\ B {\bf 638} (2006) 480
  [hep-ph/0603202].
  
\bibitem{Bernard:2009zm}
  V.~Bernard {\it et al.},
  Phys.\ Rev.\ D {\bf 80} (2009) 034034
  [arXiv:0903.1654].

\bibitem{Antonelli:2010yf}
  M.~Antonelli {\it et al.},
  Eur.\ Phys.\ J.\ C {\bf 69} (2010) 399
  [arXiv:1005.2323].

\bibitem{Callan:1966hu}
  C.~G.~Callan and S.~B.~Treiman,
  Phys.\ Rev.\ Lett.\  {\bf 16} (1966) 153.
 
\bibitem{Becirevic:2004ya}
  D.~Becirevic {\it et al.},
  Nucl.\ Phys.\  B {\bf 705} (2005) 339
  [arXiv:hep-ph/0403217].
 
 \bibitem{Carrasco:2014pta}
  L.~Riggio {\it et al.} [ETM Coll.],
  PoS LATTICE {\bf 2014} (2014) 387
  [arXiv:1411.1201 [hep-lat]].

\bibitem{Lubicz:2010bv}
  V.~Lubicz {\it et al.}  [ETM Coll.],
  PoS LATTICE {\bf 2010} (2010) 316
  [arXiv:1012.3573].
 
 \bibitem{Gasser:1984ux}
  J.~Gasser and H.~Leutwyler,
  Nucl.\ Phys.\ B {\bf 250} (1985) 517.

\bibitem{Gasser:1984gg}
  J.~Gasser and H.~Leutwyler,
  Nucl.\ Phys.\ B {\bf 250} (1985) 465.

\bibitem{Flynn:2008tg}
  J.~M.~Flynn {\it et al.}  [RBC and UKQCD Coll.],
  Nucl.\ Phys.\ B {\bf 812} (2009) 64
  [arXiv:0809.1229].

\bibitem{Bourrely:2008za} 
  C.~Bourrely, I.~Caprini and L.~Lellouch,
  Phys.\ Rev.\ D {\bf 79}, 013008 (2009)
  [Erratum-ibid.\ D {\bf 82}, 099902 (2010)]
  [arXiv:0807.2722].
   
\bibitem{Ademollo:1964sr}
  M.~Ademollo and R.~Gatto,
  Phys.\ Rev.\ Lett.\  {\bf 13} (1964) 264.

\bibitem{ETMC_Kl3}
 N.~Carrasco {\it et al.}  [ETM Coll.], paper in preparation.

\bibitem{PDG}
  K.~A.~Olive {\it et al.}  [Particle Data Group Coll.],
  Chin.\ Phys.\ C {\bf 38} (2014) 090001.
  
\end{thebibliography}
\end{document}